\documentclass[twocolumn,showpacs,preprintnumbers]{revtex4} 
\usepackage{amssymb}
\usepackage[dvips]{graphicx}

\begin{document}

%\title{What does intrinsic tunnelling spectroscopy really examine?}
\title{Reply to comment on "Essence of intrinsic tunnelling: Distinguishing intrinsic features from artefacts"}

\author{V.N. Zavaritsky$^{1,2}$}
\address
{$^{1}$Department of Physics, Loughborough University, Loughborough, United Kingdom, 
$^{2}$Kapitza Physics Institute \& General Physics Institute, Moscow, Russia\\
}

\begin{abstract}
The recent PRB, henceforth referred as Ref.[1], experimentally resolves the intrinsic shape of the c-axis current-voltage characteristics (IVC) of HTSC and demonstrates that at sufficiently high heat loads the heating-induced IVC nonlinearities exceed the intrinsic ones so radically that the latter might be safely ignored.

The author of the comment fails to take account of the experimental findings by Ref.[1] and seeks to cast doubt on all its conclusions through reference to a brush-like IVC, which is claimed to be free of heating. I will show that this claim lacks substantiation; indeed it can be stated with certainty that the IVC is not free from heating.  %I will further show that the data selected for this comment make it possible to demonstrate for the first time that the parameter-free description by Ref.[1] superimposes the vastly different IVC taken at bath temperatures spanned over 180K onto a single curve which correlates reasonably with $R(T)$ of the same sample. 
I will further show that the data selected for this comment make it possible to explore for the first time the effect of temperature on a range of loads where the genuine response is not hidden by heating and to demonstrate for the first time that $R(T)$ of the same sample is responsible for a rich variety of IVC behaviours taken above and below $T_c$ at bath temperatures spanned over 180K. Thus these data in fact provide strong novel evidence in favour of the major conclusions by Ref.[1], in particular the extrinsic cause of the key findings by intrinsic tunnelling spectroscopy.

\end{abstract}

\maketitle

Heat $W$, dissipated in a sample, escapes through its surface area $A$ and causes significant heating if the heat load $P=W/A$ exceeds the critical value $P_c$, which depends on the experimental environment. Notably, $P_c$ depends on the coolant medium; it is close to $1W/cm^2$ for liquid helium and is significantly smaller for helium vapour at a comparable temperature. Heating is probably the most common problem in low temperature research and a particularly harsh limiting factor for the study of current-voltage characteristics (IVC). Self-heating of superconductors is particularly well studied; notably, it is known to cause IVC nonlinearities and transform a single-valued IVC into a multi-valued characteristic with regularly spaced branches (see Ref.\cite{2} for a comprehensive review). 

The findings by Ref.\cite{2} are particularly relevant to high temperature superconductors (HTSC) because the exceptionally poor thermal and electrical conductivities of HTSC makes them particularly prone to local heating.  However, unlike other studies of HTSC, the heating issues in `intrinsic tunnelling' devoted to the brush-like IVC were misinterpreted or ignored until recently. Particularly confusing claims arise from `intrinsic tunnelling spectroscopy' (IJT), which postulates (i) that HTSCs {\it factually} represent natural stacks of atomic-scale intrinsic superconductor-insulator-superconductor (SIS) Josephson junctions and (ii) an intrinsic cause for the IVC features built by the heat loads in excess of { kilowatts per cm$^2$} which exceed the corresponding $P_c$ by 4-6 orders, (Refs.\cite{1,3}). 

Central to the resolution of this confusion are the systematic experimental studies summarised in Ref.\cite{1}, which  suggest that the true IVC is Ohmic above $T_c$ while the brush-like part is reasonably described by: 
\begin{equation}
|V_\#|=R_\#|(I-I^*)|;
\end{equation}                     
where the differential resistance of a resistive branch $R_\#$ is proportional to its number, $\#$, and represents a fraction of the normal state resistance $R_N(T_B)$ of the same sample measured under conditions of complete suppression of its superconductivity.  %$R_N$ depends on temperature \cite{yasuda,me} so at $P>P_c$ the branches deviate from the linearity.  
The behaviour in Eq(1) is compatible with Josephson-based explanations albeit ruling out the basic IJT postulate, see Ref.\cite{1}.

Heating masks the genuine response, which could be seen at $P<P_c$ only. Furthermore, the experiments summarised by Ref.\cite{1} show that at sufficiently high heat loads the heating-induced IVC nonlinearities exceed the intrinsic ones so radically that the latter might be safely ignored. The experimental IVC in such circumstances is primarily determined by $R_N(T)$, while the mean temperature, $T$,  of the self heated sample is appropriately described by Newton's Law of Cooling (1701),

\begin{equation}
T=T_B+P/h, 
\end{equation}
where $T_B$ is the temperature of the coolant medium (liquid or gas) and $h$ is the heat transfer coefficient, which depends neither on $A$ nor $T$, see Refs.\cite{1,5} for details. The consistency of this parameter-free description (which suggests the extrinsic cause of the key IJT findings, see Refs.\cite{1,4,5,6}) was reaffirmed by independent measurements by Ref.\cite{7}. The area independence of heating effects observed by Ref.\cite{1} was strongly supported by Ref.\cite{4}, which addressed the heating cause of IJT spectra by \cite{10} and discovered that practically the same heat loads $P\sim10kW/cm^2$ build the IJT gap in Bi2212 structures of vastly different area $1<A<30\mu m^2$. 

As is shown below, the data selected for the comment provide novel evidence in favour of the major conclusions by Ref.[1] and allow resolution of important issues which were not covered by Ref.[1]. In particular, I will provide a first demonstration of the fact that (i) the model by Ref.[1] describes quantitatively the whole set of IJT IVC taken at $T_B$ above and below $T_c$; (ii) the genuine parts of these IVC agree reasonably with Eq.(1) (iii) $P_c$ drops with temperature so radically that at low T the extrinsic features dominate almost throughout the range of loads. 

The set of drastically different IVC of the same `mesa' at various $T_B$ (promoted by Ref.\cite{9}, selected by Krasnov for the present comment, Ref.\cite{8} 
 and, for some reason, omitted from its latest, fourth revision) provides a harsh consistency check for the parameter-free description by Ref.[1] and hence is particularly pertinent to the subject under discussion. Besides, this set is worth considering as it allows exploration for the first time of the effect of temperature on a range of loads where the genuine response is not hidden by heating. To allow intrinsic features to be distinguished from extrinsic ones, it is appropriate to consider $R=V/I$ as a function of the heat load, $P = V I/A$, rather than I-V only (see above and also Refs.[1,3]). This is illustrated in Fig.1a, which shows the set of accordingly re-plotted IVC mentioned above.  As seen in Fig.1(a), all curves exhibit a similar shape: there is a well defined threshold level, $P_c$, below which $R(P)$ is flat, while it drops rapidly at $P>P_c$. According to Refs.\cite{1,5,6}, the $R(P)$ curves in the latter case are caused by Joule self-heating and hence must obey Eq.(2).  Indeed, as is seen from Fig.1(b), the parameter-free Eq.(2) collapses all IVC, obtained at $T_B$ spanned over 180K, into a single curve which reproduces quantitatively the $R(T)$ of the same 'mesa' and allows an estimate of the heat transfer coefficient $h=32Wcm^{-2}K^{-1}$, typical for this type of measurements. Thus, Fig.1 confirms the heating origin of the IVC non-linearity and suggests that the IVC by Refs.\cite{8,9},  will be almost linear above and below $T_c=93K$ if the heating artefacts are removed. So these findings suggest that Eqs.(1,2) correctly describe both the IJT spectra and a genuine IVC hidden by heating artefacts, hence demonstrating the unfoundedness of the principal claim by the author of the comment.

\begin{figure}
\begin{center}
\includegraphics[angle=-0,width=0.47\textwidth]{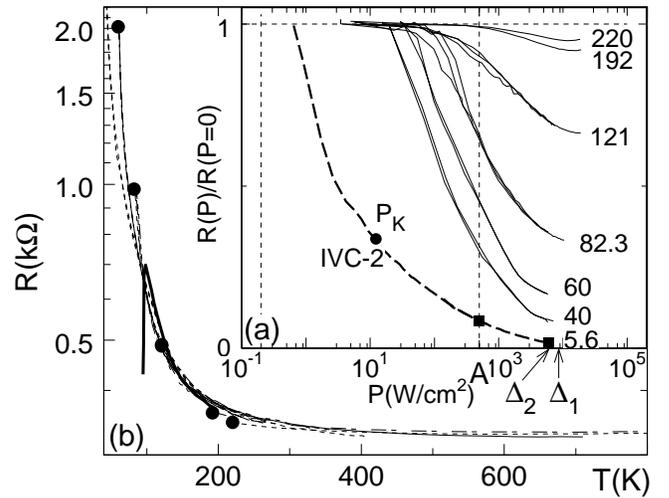}
\vskip -0.1mm
\caption{{\bf (a)}:  The solid lines represent the nonlinear IVC, measured by Ref.\cite{9} at different $T_B$ above and below $T_c=93K$, re-plotted as a sample resistance, $R$=$V/I$, normalised by its value at $P$$\rightarrow$$0$, versus the heat load, $P$=$IV/A$; $A$=$26\mu m^2$.  $T_B$ are shown in the figure at the corresponding curves; the solid dots in Fig.1(b) represent $R(P$$\rightarrow$$0)$ vs $T_B$. The thick broken line shows $R(P)$ for \mbox{IVC-2}; the characteristic heat loads which build the IJT gaps and the point $'A'$ in the comment's IVC are shown by the solid dots and the axis labels. $P_K$ marks the typical heat load of a domestic kettle, which is 2-3 orders smaller than P which builds the `gaps' advocated by the author of the comment.\\
{\bf (b)}: Compares the measured $R(T)$ shown by the thick solid line with the ones calculated with Eq.(1) from the nonlinear IVC using one and the same heat transfer coefficient  $h=32Wcm^{-2}K^{-1}$ for the data taken at $T_B$ spanned over 180K.   
}
\end{center}
\end{figure}

Fig.1(b) strongly suggests that $h$ does not depend on temperature. This conclusion agree well with the earlier experimental findings summarized by Ref.\cite{1} and similar, albeit less reliable, conclusions could be drawn from the row data by Ref.\cite{7}. However, it contradicts radically the findings by Ref.\cite{10}. A resolution of this dichotomy is possible by taking into account the actual experimental arrangement of Ref.\cite{10} where the thermometer and the overheated sample are individually heat-sinked to the bath through metal electrodes of enhanced area and thermal conductivity. Such an arrangement makes unavoidable a thermal lag between the thermometer and the overheated sample. This lag depends on $P$ and $T_B$ and, for this reason, the data by \cite{10} are not beyond dispute, see ref.\cite{4} for more details.

As is seen from Fig.1a, the critical load drops with temperature radically, as does the range of loads where intrinsic features dominate. However, the heat loads of the characteristic IVC points demonstrate the opposite trend, as both the switching current $I_J$ and IJT gap increase as the temperature becomes lower. So a study of intrinsic response, feasible at temperatures slightly below $T_c$ and above (see Ref.\cite{1}), becomes enormously complicated at helium temperatures, where the extrinsic features dominate almost throughout the range of loads where the brush-like IVC exists. This common case is illustrated by the appropriately re-plotted \mbox{IVC-2} ($T_B=5.6K$) from Fig.2 of the %latest (fourth) revision of the 
comment. As is seen in Fig.1(a), the entire \mbox{IVC-2} belongs to the falling part of $R(P)$ and so is most likely caused by heating. This conclusion is additionally supported by (i) the qualitative similarity of the falling parts of $R(P)$ taken at vastly different $T_B$ and (ii) the reasonable correlation between the coarse estimate of $P_c(5.6K)$ for \mbox{IVC-2}, the value (shown by the unlabelled grid line in Fig.1(a)) obtained from the measurements of the local heating in another 'mesa' (see Refs.\cite{1,3} for details) and the extrapolation of $P_c(T_B)$ from Fig.1(a).  

Thus, the \mbox{IVC-2} supports neither the claim that "the self-heating along the branches is negligible" nor the claim that "the genuine interlayer IVC's are strongly nonlinear".  The last major claim of the comment, that the branches in the brush "are perfectly periodic" is also at odds with the experiment because neither the genuine branches described by Eq(1), nor the nonlinear ones, advocated by the author of the comment, obey the definition of periodicity:
\begin{equation}
F(x+a)=F(x), \hspace{3mm} a=const. 
\end{equation}

%As the variety of the experimental $R_N(T)$ behaviours (discussed in Ref.[1])is sufficient to generate all known shapes of the IJT gap-like feature, it would be fallacious to associate the  IVC `back-bending' with "the case of extreme self-heating". This conclusion is consistent with the earlier studies which suggest the extrinsic cause of all known shapes of the IJT gaps hence rendering irrelevant the association of the IVC `back-bending' with the signature of "extreme self-heating", see Ref.[1] for details. 
                            
As far as the heating in the samples of different $A$ is concerned, neither the critical current $I_c$ nor the heat $W=IV$ (confusingly denoted as $P$ in the comment) fit the comparison (see above). Furthermore, $I_J$ depends on ambient factors (eg. it is easily suppressed by a small magnetic field) which leave unaffected both the overall shape of IVC and the IJT gap. For this reason and because of the unknown cause of $I_J$, it is more appropriate to compare the heat loads $\Delta$ which build the IJT gap. $A$ and $\Delta$ are estimated accordingly: $A_1\cong60\mu m^2$ and $\Delta_1\cong8.6kW/cm^2$ are taken from the figure assuming that the author alleges $I_J$ with $I_c$; $A_2=13.5\mu m^2$ and $\Delta_2\cong6kW/cm^2$ are taken from the source article Ref.[10]. As $\Delta_1$ and $\Delta_2$ are practically the same (see Fig.1a), there are no valid reasons to expect the self-heating to be radically different in these samples, one of which is declared to represent `the case of extreme self-heating' by the author of the comment, see Ref.\cite{11}. This conclusion is consistent with the earlier studies which suggest the extrinsic cause of all known shapes of IJT gaps, hence rendering irrelevant the association of the IVC `back-bending' with the signature of extreme self-heating, see Ref.[1] for details. 

To conclude, it is demonstrated, using exclusively the data selected for the present comment by its author, that neither the argumentation nor the conclusions of the comment by V. Krasnov are borne out by experiment. Contrary to the comment's claims, Ref.\cite{1} addresses the genuine IVC experimentally and shows that at sufficiently high heat loads the heating-induced IVC nonlinearities exceed the plausible intrinsic ones, eg. of Eq.(1), so radically that the latter might be safely ignored. Moreover, the data selected for this comment make it possible to explore for the first time the effect of temperature on the range of loads where the genuine response is not hidden by heating. Furthermore, it is shown for the first time that the whole set of experimental IVC taken above and below $T_c$ at vastly different $T_B$ spanned over 180K are described quantitatively by Newton's Law of Cooling and Ohm's law using the normal state resistance of the same sample only. This novel finding confirms the heating origin of the IVC non-linearity which was originally claimed as "Evidence for Coexistence of the Superconducting Gap and the Pseudogap" in Ref.\cite{9} and suggests that unlike conventional spectroscopy \cite{STM}, the heating in IJT is not a small perturbation but a principal cause of IVC nonlinearity. 

Our conclusions do not rule out worthwhile IJT experiments, some of which were proposed by Ref.\cite{1}. Moreover, the feasibility of macroscopic quantum tunnelling (MQT) was recently discussed by the authors of Refs.\cite{12,13}. Such studies might be virtually unaffected by heating as long as they appropriately address the statistics of stochastic switching from a zero-$P$ state. However, heating can spoil MQT, and indeed the authors of Ref.\cite{14} discovered that the escape process from the first resistive branch is most likely governed by heating even in the bridge-like samples which reveal a noticeably higher $P_c$ than the 'mesas' considered above, see Ref.\cite{1}. The findings by Ref.\cite{14} thus provide independent evidence in support of our conclusions. However, the range of heat loads where MQT still exists remains to be explored eg., by an in-situ suppression of the switching current by magnetic field. 

\centerline{\bf Acknowledgement}

I am grateful to the %first referee for drawing my attention to Refs.\cite{12,13} and the 
authors of Ref.\cite{13} for their explanation of their belief that IVC of HTSC bridge taken at millikelvin temperatures remains unaffected by heating even at P$\sim$1kW/cm$^2$. The author wishes to record his strong objection to the manner in which the reply has been edited.


\begin{thebibliography}{99}
\bibitem{8} see V. M. Krasnov, cond-mat/0601101 for the original version of the Comment submitted to PRB (cond-mat/0601101v1) and its modifications in response to my criticism; note that the data for $T_B=$192 and 220K as well as the very high-P data (for $T_B>$40K) in Fig.1 are drawn from this source. %This reference is removed following the editorial demand which is nothing less than an ultimatum. So, I consider that it was removed by the editor, not the author.
\bibitem{11} Detailed comparison of the actual overheating in these samples is not possible in the absence of reliable $R_N(T)$ or $h$.
\end{thebibliography}
\end{document}